\documentclass{jpsj3}
\usepackage{bm}
\usepackage{color}
\title{%
  Surface Resistance and Amplitude Mode
  under Uniform and Static External Field
  in Conventional Superconductors
}
\author{%
  Takanobu Jujo\thanks{E-mail address: jujo@ms.aist-nara.ac.jp}
}
\inst{%
  Division of Materials Science,
  Graduate School of Science and Technology,
  Nara Institute of Science and Technology,
  Ikoma, Nara 630-0101, Japan
}
\abst{%
  We calculate the surface resistance of s-wave superconductors
  under a static external field 
  on the basis of the quasiclassical approximation.
  The numerical calculations are performed both for the dirty and
  relatively clean cases,
  and the difference 
  in the absorption spectrum between them is investigated.
  The amplitude mode in the dirty case,
  which has been previously studied in the local response function, 
  appears also in the nonlocal response.
  In the clean case,
  there is a large discrepancy between 
  the excitation energy of the amplitude mode
  and the energy of the gap edge
  where the coupling between the electrons and the external field
  is effective. 
  Therefore, the amplitude mode exists
  even when the superconductor is relatively clean,
  but its contribution to the response function is small.
  These behaviors, which are qualitatively consistent with
  the experimental results, 
  are obtained by taking into account
  the static field nonperturbatively.
}
\begin{document}
\setlength{\textwidth}{504pt}
\setlength{\columnsep}{14pt}
\hoffset-23.5pt
\maketitle

  \section{Introduction}

In recent years,
the observation of nonequilibrium states
in the superconducting state
has progressed owing to the advancement
of the microwave spectroscopy in experiments.~\cite{giannetti}
The excitation and relaxation
of quasiparticles and collective modes
have been observed by pump-probe spectroscopy.~\cite{beck,matsunaga}
As a major development, the amplitude mode,
which was difficult to observe
because it did not appear
in the linear response,
has been discussed in the nonlinear response
with experimental~\cite{matsunaga14,katsumiprl,isoyama,grasset}
and theoretical methods.~\cite{tsujiaoki,jujo17,murotani,silaev,schwarzhaenel}

On the other hand, the generation of nonequilibrium states
is possible not only by optical excitation
but also by stationary fields (magnetic field and current),
which has been studied for a long time.
For example, one-particle properties
such as the density of states have been investigated
experimentally~\cite{levine,anthore} and theoretically~\cite{fulde}.
A similarity of the one-particle spectrum
between dirty superconductors under a static field
and those including paramagnetic impurities
was discussed.~\cite{makifulde}
Collective modes are 
included in the one-particle state 
through self-energy,~\cite{jujo19}
but these characteristics are obscured through
integrations by the energy and wave numbers.
On the other hand,
since in the response function 
collective modes can emerge through
the final state interaction (vertex corrections),
it is possible that
these characteristics appear more directly.

The response function in the nonequilibrium state
under a static field has also been studied
theoretically,~\cite{maki65,ovchinnikov,gurevich,kubo,moor}
but there remain several properties to be 
investigated as compared with the case of the one-particle state.
For example, in the surface resistance,
the existence (absence) of a peak structure around
the gap edge has been experimentally observed
for dirty (clean) superconductors.~\cite{budzinski,budzinskiB}
A theoretical proposal with a surface state
has been made to solve this problem,~\cite{pincus,sherman}
but its relevance to the response function is not obvious.
The perturbative calculation with static fields 
suggested a relationship between
the peak structure in the response function
and the collective mode.~\cite{ovchinnikov78,moor}
(The peak has also been observed recently in an experiment
in dirty $s$-wave superconductors.~\cite{nakamura})
However, the perturbative calculation cannot explain
the absorption edge and the absence of the peak in 
the clean case.

The gap edge of the density of states
under a static field
is not determined by the superconducting order parameter
irrespective of whether the superconductor is dirty or clean.
Therefore, in order to calculate the surface resistance,
it is necessary to take into account the static field 
nonperturbatively.
In this paper, we consider this problem
and calculate the nonlocal response function
under a static field to derive the surface resistance 
in both dirty and relatively clean superconductors.
So far, the amplitude mode has been discussed
on the basis of the local response function. 
In this study, we show that the 
amplitude mode can appear even when nonlocality is important,
and we clarify the reason why the amplitude mode is not visible
in the absorption spectrum in the case of relatively
clean superconductors.

The structure of this paper is as follows.
Section 2 shows how to calculate the surface impedance
using the quasiclassical approximation.
This calculation derives vertex correction terms
due to impurity scattering
and electron--phonon interaction.
The latter includes the amplitude mode. 
Section 3 shows the results of the numerical calculation
of the surface resistance,
and the relationship between
the one-particle spectrum and the response function
is discussed.
In Sect. 4, the relationship between
the calculated and
experimental results is discussed.
We put $\hbar=c=1$ in this paper ($c$ is the velocity of light).

\section{Formulation}

  The surface impedance at the frequency $\omega$ is given by~\cite{reuter} 
  \begin{equation}
    Z_S=\frac{4\pi E_{\omega}^x(z=0)}{H_{\omega}^y(z=0)}.
  \end{equation}
  We consider the Cartesian coordinate system $(x,y,z)$. 
  The external field is in the $xy$-plane, and there is
  an interface at $z=0$
  between the superconductor ($z > 0$) and the vacuum ($z<0$).
  The electric field $E_{\omega}^x(z)$ [the magnetic field
    $H_{\omega}^y(z)$], which
  is parallel to the $x\text{ $(y)$ }$-axis and penetrates into $z>0$
  in the superconductor, 
  is determined by the
  Maxwell equation
  \begin{equation}
      \frac{dE_{\omega}^x(z)}{dz}=i\omega H_{\omega}^y(z)     \text{  and  }      -\frac{dH_{\omega}^y(z)}{dz}=4\pi j_{\omega}^x(z).      
    \label{eq:zmaxwelleq}
  \end{equation}
  $j_{\omega}^x(z)$ is the current.
  (The displacement current term is omitted because of its smallness
  as mentioned in Sect. 3.)
  Using the Fourier transformation with
  the specular boundary condition~\cite{abrikosov}
  [$E^x_{\omega}(z)=E^x_{\omega}(-z)$ and
  $j_{\omega}^x(z)=j_{\omega}^x(-z)$]
  : $\tilde{E}_{\omega}^x(q)
  =\int dze^{-iqz}E^x_{\omega}(z)$
  and $\tilde{j}_{\omega}^x(q)=\int dze^{-iqz}j_{\omega}^x(z)$,
  Eq. (\ref{eq:zmaxwelleq}) is written as
    \begin{equation}
    -q^2\tilde{E}_{\omega}^x(q)
    =2E^{x'}_{\omega}(z=0)
    -4\pi i\omega \tilde{j}_{\omega}^x(q)
  \end{equation}
    with   $E^{x'}_{\omega}(z=0)
=\left.dE_{\omega}^x(z)/dz\right|_{z=0}
    =4\pi i\omega\int_0^{\infty}j_{\omega}^x(z)dz$.
Then, the surface impedance is rewritten as 
  \begin{equation}
    Z_S=\frac{4\pi i\omega E^x_{\omega}(0)}
    {E^{x'}_{\omega}(0)}=
    4\pi i\omega\int\frac{dq}{2\pi}
    \frac{-2}
         {q^2-4\pi i\tilde{j}_{\omega}^x(q)/\tilde{E}^x_{\omega}(q)}.
         \label{eq:surfimp}
  \end{equation}

  The current $j_{\omega}^x(z)$ is given by using
  the nonequilibrium quasiclassical Green function
  as~\cite{eliashberg}
  \begin{equation}
    j_{\omega}^x(z)=
    e\int_{FS}v_{\mib k}^x\frac{mk_F}{2\pi}\int\frac{d\epsilon}{4\pi i}
    Tr[T_{\epsilon^-}^h
      \hat{g}^{+'}_{{\mib k},\epsilon^+,\epsilon^-}(z)-
      T_{\epsilon^+}^h
      \hat{g}^{-'}_{{\mib k},\epsilon^+,\epsilon^-}(z)
      +\hat{g}^{(a)'}_{{\mib k},\epsilon^+,\epsilon^-}(z)].
    \label{eq:current1}
  \end{equation}
  Here, $m$ and $k_F$ are the mass of quasiparticles and
  the Fermi wave number, respectively.
  $v_{\mib k}^x$ is the quasiparticle
  velocity parallel to the $x$-axis [${\mib v}_{\mib k}=(v^x_{\mib k},
    v^y_{\mib k},v^z_{\mib k})$] at
  the Fermi surface ($|{\mib k}|=k_F$ and $|{\mib v}_{\mib k}|=v_F$).
    $\epsilon^{\pm}=\epsilon\pm\omega/2$ and  
  $T^h_{\epsilon}:={\rm tanh}(\epsilon/2T)$
  ($T$ is the temperature).
    $\int_{FS}$ and $Tr$ mean
  the integration over the Fermi surface and
  taking the trace of the matrix, respectively.
    The nonequilibrium
    quasiclassical Green function
    [$\hat{g}^{\pm'}_{{\mib k},\epsilon^+,\epsilon^-}(z)$ and
      $\hat{g}^{(a)'}_{{\mib k},\epsilon^+,\epsilon^-}(z)$]
    under a uniform and static external field
  (represented by a vector potential ${\mib A}_0$)
  with the first order of $A^x_{\omega}(z)=E_{\omega}^x(z)/i\omega$
 satisfies
  the following equations:~\cite{eliashberg}
  \begin{equation}
    \begin{split}
      &  \hat{\tau}_3
      \left[\epsilon^+\hat{1}
        +e{\mib v}_{\mib k}\cdot{\mib A}_0\hat{1}
        -\hat{\Sigma}_{\epsilon^+}^{+}\right]
      \hat{g}^{(a)'}_{\mib k,\epsilon^+,\epsilon^-}(z)
      -\hat{g}^{(a)'}_{\mib k,\epsilon^+,\epsilon^-}(z)
      \left[\epsilon^-\hat{1}
        +e{\mib v}_{\mib k}\cdot{\mib A}_0\hat{1}
        -\hat{\Sigma}_{\epsilon^-}^-\right]\hat{\tau}_3
      +iv^z_{\mib k}
\frac{d \hat{g}^{(a)'}_{\mib k,\epsilon^+,\epsilon^-}(z)}{dz}\\
      &  -\hat{\tau}_3\hat{\Sigma}^{(a)'}_{\epsilon^+,\epsilon^-}(z)
      \hat{g}^-_{{\mib k},\epsilon^-}
 +\hat{g}^+_{{\mib k},\epsilon^+}\hat{\Sigma}^{(a)'}_{\epsilon^+,\epsilon^-}(z)
      \hat{\tau}_3
      -\left(T^h_{\epsilon^+}-T^h_{\epsilon^-}\right)ev^x_{\mib k}
      A^x_{\omega}(z)
      \left(\hat{g}^+_{{\mib k},\epsilon^+}\hat{\tau}_3
      -\hat{\tau}_3\hat{g}^-_{{\mib k},\epsilon^-}\right)=0
    \end{split}
    \label{eq:kineqanomal}
  \end{equation}
      and
  \begin{equation}
    \begin{split}
      &  \hat{\tau}_3
      \left[\epsilon^+\hat{1}
        +e{\mib v}_{\mib k}\cdot{\mib A}_0\hat{1}
        -\hat{\Sigma}_{\epsilon^+}^+\right]
      \hat{g}^{+'}_{\mib k,\epsilon^+,\epsilon^-}(z)
      -\hat{g}^{+'}_{\mib k,\epsilon^+,\epsilon^-}(z)
      \left[\epsilon^-\hat{1}
        +e{\mib v}_{\mib k}\cdot{\mib A}_0\hat{1}
        -\hat{\Sigma}_{\epsilon^+}^-\right]\hat{\tau}_3
      +iv^z_{\mib k}\frac{d
      \hat{g}^{+'}_{\mib k,\epsilon^+,\epsilon^-}(z)}{dz}\\
      &  -\hat{\tau}_3\hat{\Sigma}^{+'}_{\epsilon^+,\epsilon^-}(z)
      \hat{g}^+_{{\mib k},\epsilon^-}
      +\hat{g}^+_{{\mib k},\epsilon^+}
      \hat{\Sigma}^{+'}_{\epsilon^+,\epsilon^-}(z)
      \hat{\tau}_3
      -e v^x_{\mib k}
      A^x_{\omega}(z)
      \left(\hat{g}^+_{{\mib k},\epsilon^+}\hat{\tau}_3
      -\hat{\tau}_3\hat{g}^+_{{\mib k},\epsilon^-}\right)=0,
    \end{split}
    \label{eq:kineqretadv}
  \end{equation}
  where
  $\hat{1}=\left(\begin{smallmatrix} 1 & 0 \\0& 1\end{smallmatrix}\right)$
    and
    $\hat{\tau}_3
    =\left(\begin{smallmatrix} 1 & 0 \\0& -1\end{smallmatrix}\right)$.
      $\hat{g}^{\pm}_{{\mib k},\epsilon}$ and $\hat{\Sigma}^{\pm}_{\epsilon}$
      are
      the one-particle Green function and
      self-energy under the uniform and static field, respectively
      [$+$ $(-)$ corresponds to
        the retarded (advanced) function].
      $\hat{g}^{\pm}_{{\mib k},\epsilon}$ 
      depends on the direction of Fermi wave number ${\mib k}$
      because of the term
      $e{\mib v}_{\mib k}\cdot{\mib A}_0$.
          $\hat{\Sigma}^{\pm'}_{\epsilon^+,\epsilon^-}(z)$
    and
    $\hat{\Sigma}^{(a)'}_{\epsilon^+,\epsilon^-}(z)$
    are the self-energies in the nonequilibrium state, which include
    the effect of $E_{\omega}^x(z)$ in the first order.
      We consider the electron--phonon interaction
  within the weak-coupling approximation~\cite{note1} and
  the impurity scattering with Born approximation.
  Then, the self-energy is written as
  \begin{equation}
    \hat{\Sigma}^{\pm}_{\epsilon}=
   \Delta\hat{\tau}_1+\alpha
    \int_{FS}\hat{\tau}_3\hat{g}^{\pm}_{{\mib k},\epsilon}
    \hat{\tau}_3.
    \label{eq:equiliselfenergy}
  \end{equation}
 Here, $\hat{\tau}_1=
  \left(\begin{smallmatrix} 0 & 1 \\1& 0\end{smallmatrix}\right)$
    and $\alpha:=(mk_F/2\pi)n_iu_i^2$, where
    $n_i$ and $u_i$ are the concentration and potential of nonmagnetic
    impurities, respectively.
    [$\hat{g}^{\pm}_{{\mib k},\epsilon}$ is given in
      Eq. (\ref{eq:onepartg}).]
    The superconducting gap $\Delta$ is determined by the
    gap equation
    \begin{equation}
      \Delta=p\int\frac{d\epsilon}{2\pi i}T_{\epsilon}^h
      \int_{FS}
      \frac{-1}{2}Tr[\hat{\tau}_1
        (\hat{g}^{+}_{{\mib k},\epsilon}-
        \hat{g}^{-}_{{\mib k},\epsilon})].
      \label{eq:gapeq}
    \end{equation}
    Here, $p:=(mk_F/2\pi)(g_{ph}^2/\omega_D)$, where
    $g_{ph}$ is the electron-phonon coupling constant and
    $\omega_D$ the Debye frequency.
    Corresponding to this approximation,
      the self-energy in the nonequilibrium state is given by
$\hat{\Sigma}^{\pm'}_{\epsilon^+,\epsilon^-}(z)
      =\hat{\Sigma}^{(ep)}_{\omega}(z)
      +\hat{\Sigma}^{\pm(im)}_{\epsilon^+,\epsilon^-}(z)$
        with
      \begin{equation}
      \hat{\Sigma}^{(ep)}_{\omega}(z)=
      p\int\frac{d\epsilon}{2\pi i}\int_{FS}\hat{\tau}_3
      \left[
        T^h_{\epsilon^-}\hat{g}^{+'}_{\mib k,\epsilon^+,\epsilon^-}(z)
        -T^h_{\epsilon^+}\hat{g}^{-'}_{\mib k,\epsilon^+,\epsilon^-}(z)
        +\hat{g}^{(a)'}_{\mib k,\epsilon^+,\epsilon^-}(z)
        \right]\hat{\tau}_3
      \label{eq:zepsignoneq}
      \end{equation}
      and
      \begin{equation}
      \hat{\Sigma}^{\pm(im)}_{\epsilon,\epsilon'}(z)=
      \alpha\int_{FS}\hat{\tau}_3
      \hat{g}^{\pm'}_{\mib k,\epsilon,\epsilon'}(z)\hat{\tau}_3.
  \end{equation}
  \begin{equation}
    \begin{split}
      \hat{\Sigma}^{(a)'}_{\epsilon^+,\epsilon^-}(\mib r)=&
      (T^h_{\epsilon^+}-T^h_{\epsilon^-})      \hat{\Sigma}^{(ep)}_{\omega}(z)
    +
      \alpha\int_{FS}\hat{\tau}_3
      \hat{g}^{(a)'}_{\mib k,\epsilon^+,\epsilon^-}(z)\hat{\tau}_3.
    \end{split}
  \end{equation}
      [In the weak-coupling approximation,
      the diagonal term of $\hat{\Sigma}^{(ep)}_{\omega}(z)$
      vanishes because of the relation
      $\hat{g}^{+}_{\mib k,\epsilon^{\pm}}\leftrightarrow
      -\hat{\tau}_3\hat{g}^{-}_{\mib k,\epsilon^{\mp}}\hat{\tau}_3$
      under $\epsilon\leftrightarrow -\epsilon$
      and $\mib k\leftrightarrow -\mib k$
      in the integration of
      Eq. (\ref{eq:zepsignoneq}), and 
      only the off-diagonal term remains in
      $\hat{\Sigma}^{(ep)}_{\omega}(z)$.]
    The one-particle Green function under the uniform and  static field
  is given by
  \begin{equation}
    \hat{g}^{\pm}_{{\mib k},\epsilon}=-\frac{
      \tilde{\epsilon}^{\pm}_{\mib k}\hat{1}
      +\tilde{\Delta}^{\pm}_{\epsilon}\hat{\tau}_1}
        {\sqrt{(\tilde{\Delta}^{\pm}_{\epsilon})^2
            -(\tilde{\epsilon}^{\pm}_{\mib k})^2}}.
        \label{eq:onepartg}
  \end{equation}
Here,
  $\tilde{\epsilon}^{\pm}_{\mib k}
  =\epsilon+e{\mib v}_{\mib k}\cdot{\mib A}_0-\Sigma^{\pm}_n
  (\epsilon)$
and
$\tilde{\Delta}^{\pm}_{\epsilon}=\Sigma^{\pm}_a(\epsilon)$,
in which
$\Sigma^{\pm}_n(\epsilon)\hat{1}
+\Sigma^{\pm}_a(\epsilon)\hat{\tau}_1=
\hat{\Sigma}^{\pm}_{\epsilon}$ [Eq. (\ref{eq:equiliselfenergy})].
We take the effect of 
$e{\mib v}_{\mib k}\cdot{\mib A}_0$
into account nonperturbatively
in the calculation of the conductivity.

\subsection{Vertex corrections}

We solve
Eqs. (\ref{eq:kineqretadv}) and (\ref{eq:kineqanomal}), and 
the current [the Fourier transform of Eq. (\ref{eq:current1})]
is given by 
\begin{equation}
  \tilde{j}_{\omega}^x(q)
=\tilde{j}_{\omega}^{x(0)}(q)+\tilde{j}_{\omega}^{x(im)}(q)+
\tilde{j}_{\omega}^{x(ep)}(q),
\label{eq:currentq}
\end{equation}
in which
$\tilde{j}_{\omega}^{x(0)}(q)$,
$\tilde{j}_{\omega}^{x(im)}(q)$, and
$\tilde{j}_{\omega}^{x(ep)}(q)$ indicate 
the term with no vertex corrections,
the vertex correction term by impurity scattering,
and the amplitude fluctuation mode term, respectively.
These are written as 
  \begin{equation}
    \begin{split}
      \tilde{j}_{\omega}^{x(\text{x})}(q)
      =
      \frac{-3}{8\lambda_0^2}
    \int\frac{d\epsilon}{2\pi i}
    \left[
      T^h_{\epsilon^-}\kappa^{(\text{x})++}_{\epsilon^+,\epsilon^-}(q)
      -      T^h_{\epsilon^+}\kappa^{(\text{x})--}_{\epsilon^+,\epsilon^-}(q)
      +(T^h_{\epsilon^+}-T^h_{\epsilon^-})
\kappa^{(\text{x})+-}_{\epsilon^+,\epsilon^-}(q)
\right]
             \tilde{A}_{\omega}^x(q)
    \end{split}
    \label{eq:currentkappa}
  \end{equation}
  with  $\text{x}=0$, $im$, and $ep$.
  ($\lambda_0$ is the magnetic field penetration depth and
  $1/\lambda_0^2=4\pi e^2n_e/m$ with the electron density
  $n_e=k_F^3/3\pi^2$.)
  $\kappa^{(0)ss'}_{\epsilon^+,\epsilon^-}(q)$,
  $\kappa^{(im)ss'}_{\epsilon^+,\epsilon^-}(q)$,
  and $\kappa^{(ep)ss'}_{\epsilon^+,\epsilon^-}(q)$
  are given by 
  \begin{equation}
    \kappa^{(0)ss'}_{\epsilon^+,\epsilon^-}(q)
=    \int_{FS}
    \left(\frac{v_{\mib k}^x}{v_F}\right)^2
    \frac{(\zeta^s_{{\mib k},\epsilon^+}
      +\zeta^{s'}_{{\mib k},\epsilon^-})(1+g^s_{{\mib k},\epsilon^+}
      g^{s'}_{{\mib k},\epsilon^-}
      +f^s_{{\mib k},\epsilon^+}
      f^{s'}_{{\mib k},\epsilon^-})}
         {(v^z_{\mib k}q)^2+(\zeta^s_{{\mib k},\epsilon^+}
           +\zeta^{s'}_{{\mib k},\epsilon^-})^2}
         \label{eq:defkappa0}
  \end{equation}
  ($s,s'=\pm$),
  $\zeta^{\pm}_{{\mib k},\epsilon}=
\sqrt{(\tilde{\Delta}^{\pm}_{\epsilon})^2
  -(\tilde{\epsilon}^{\pm}_{\mib k})^2}$,
and
$g^{\pm}_{{\mib k},\epsilon}\hat{1}
+f^{\pm}_{{\mib k},\epsilon}\hat{\tau}_1=
\hat{g}^{\pm}_{{\mib k},\epsilon}$.
  \begin{equation}
    \kappa^{(im)ss'}_{\epsilon^+,\epsilon^-}(q)=
    \frac{\alpha[\chi_1'^2-\chi_3'^2
        +\alpha(\chi_1'^2\chi_2-2\chi_1'\chi_3'\chi_3
        +\chi_1\chi_3'^2)]}
         {(1-\alpha \chi_1)(1+\alpha \chi_2)+(\alpha \chi_3)^2}
  \end{equation}
  and
  \begin{equation}
    \kappa^{(ep)ss'}_{\epsilon^+,\epsilon^-}(q)=
    2\tilde{\Sigma}_a(\omega,q)\kappa^{ss'(v1)}_{\epsilon^+,\epsilon^-}(q)
\label{eq:sigelph}
  \end{equation}
  with
  \begin{equation}
  \kappa^{ss'(v1)}_{\epsilon^+,\epsilon^-}(q)=
     \frac{\chi_3'+\alpha(\chi_1'\chi_3-\chi_1\chi_3')}
      {(1-\alpha \chi_1)(1+\alpha \chi_2)+(\alpha \chi_3)^2},
  \end{equation}
    \begin{equation}
    \begin{split}
    \tilde{\Sigma}_a(\omega,q)=&
    \frac{
    -(p/2)\int\frac{d\epsilon}{2\pi i}
    \left[T^h_{\epsilon^-}
    \kappa^{++(v1)}_{\epsilon^+,\epsilon^-}(q)
    -T^h_{\epsilon^+}
    \kappa^{--(v1)}_{\epsilon^+,\epsilon^-}(q)
    +(T^h_{\epsilon^+}-T^h_{\epsilon^-})
    \kappa^{+-(v1)}_{\epsilon^+,\epsilon^-}(q)
\right]    }
         {
           1+p\int\frac{d\epsilon}{2\pi i}
    \left[T^h_{\epsilon^-}
\kappa^{++(v0)}_{\epsilon^+,\epsilon^-}(q)
    -T^h_{\epsilon^+}
    \kappa^{--(v0)}_{\epsilon^+,\epsilon^-}(q)
    +(T^h_{\epsilon^+}-T^h_{\epsilon^-})
    \kappa^{+-(v0)}_{\epsilon^+,\epsilon^-}(q)
    \right]},
         \label{eq:ampelphvc}
         \end{split}
  \end{equation}
    and
    \begin{equation}
        \kappa^{ss'(v0)}_{\epsilon^+,\epsilon^-}(q)=
    \frac{\chi_2+\alpha(\chi_3^2-\chi_1\chi_2)}
         {(1-\alpha \chi_1)(1+\alpha \chi_2)+(\alpha \chi_3)^2}.
    \end{equation}
    Here,
    \begin{equation}
      \begin{pmatrix}
        \chi'_1 \\ \chi'_3 \\ \chi'_2
      \end{pmatrix}
=    \int_{FS}
      \frac{(v_{\mib k}^x/v_F)
        (\zeta^s_{{\mib k},\epsilon^+}
      +\zeta^{s'}_{{\mib k},\epsilon^-})
      }
         {(v^z_{\mib k}q)^2+(\zeta^s_{{\mib k},\epsilon^+}
           +\zeta^{s'}_{{\mib k},\epsilon^-})^2}
\begin{pmatrix}
         1+g^s_{{\mib k},\epsilon^+}g^{s'}_{{\mib k},\epsilon^-}
      +f^s_{{\mib k},\epsilon^+}
      f^{s'}_{{\mib k},\epsilon^-}
      \\
      g^s_{{\mib k},\epsilon^+}f^{s'}_{{\mib k},\epsilon^-}
      +f^s_{{\mib k},\epsilon^+}g^{s'}_{{\mib k},\epsilon^-}
      \\
      -1+g^s_{{\mib k},\epsilon^+}g^{s'}_{{\mib k},\epsilon^-}
      +f^s_{{\mib k},\epsilon^+}
      f^{s'}_{{\mib k},\epsilon^-}
\end{pmatrix},
\label{eq:defchi}
    \end{equation}
    and $\chi_{1,2,3}$ are defined by similar relations with
    $(v_{\mib k}^x/v_F)$ replaced by $1$
    in Eq. (\ref{eq:defchi}).
    $\tilde{\Sigma}_a(\omega,q)$ is related to
    the off-diagonal part of the Fourier transform of
    Eq. (\ref{eq:zepsignoneq}) as
    \begin{equation}
    \tilde{\Sigma}_a(\omega,q)=\frac{-1}{2ev_F\tilde{A}_{\omega}^x(q)}
    Tr\left[\frac{1}{2}\hat{\tau}_1
      \tilde{\hat{\Sigma}}^{(ep)}_{\omega}(q)\right].
    \end{equation}

  The vertex correction terms take different values
  depending on the direction of ${\mib A}_0$.
  The integrand of $\chi'_{1,2,3}$ is
  proportional to $v_{\mib k}^x$.
When
${\mib A}_0$ is parallel to $A^x_{\omega}$
($e{\mib v}_{\mib k}\cdot{\mib A}_0=ev_{\mib k}^x A_0$),
$\chi'_{1,2,3}\ne 0$ and 
the vertex correction terms take
finite values.
On the other hand,
when ${\mib A}_0$ is perpendicular to $A^x_{\omega}$
($e{\mib v}_{\mib k}\cdot{\mib A}_0=ev_{\mib k}^y A_0$),
$\chi'_{1,2,3}= 0$ ($\chi_{1,2,3}\ne 0$)
and then the vertex correction
terms vanish.
We can consider the calculated results of 
the term with no vertex correction
in Sect. 3 as the conductivity in the case that 
${\mib A}_0$ is perpendicular to $A^x_{\omega}$.

\subsection{Absorption below $\omega<2\Delta$}

In this subsection, we show that
there is a finite absorption below $\omega<2\Delta$
even if $\alpha\ll \Delta$ when the self-energy correction
is taken into account.
For example, the real part of the optical conductivity
with no vertex correction is written as
\begin{equation}
  \begin{split}
    &    {\rm Re}\left(\frac{\tilde{j}_{\omega}^{x(0)}(q)}
    {\tilde{E}_{\omega}^{x}(q)}\right)
=
\frac{3}{2\pi\omega\lambda_0^2}
\int_0^{\omega/2} d\epsilon
    {\rm Re} \left[   \kappa^{(0)+-}_{\epsilon^+,\epsilon^-}(q)
      -\kappa^{(0)++}_{\epsilon^+,\epsilon^-}(q)
    \right]
  \end{split}
  \label{eq:absbelow}
\end{equation}
for $T=0$ and $\omega>0$.
$\kappa^{(0)ss'}_{\epsilon^+,\epsilon^-}(q)$
is given in Eq. (\ref{eq:defkappa0}).
In the limit of small $\alpha$,
$\zeta^{\pm}_{{\mib k},\epsilon}$
is given by
\begin{equation}
  \zeta^{\pm}_{{\mib k},\epsilon}\simeq
  \zeta^{'\pm}_{{\mib k},\epsilon}+
  \alpha\zeta^{'\pm}_{{\mib k},\epsilon}
  \int_{FS}(1/\zeta^{'\pm}_{{\mib k},\epsilon})
  \label{eq:approxzeta}
\end{equation}
with
\begin{equation}
\zeta_{{\mib k},\epsilon}^{'\pm}=
\sqrt{\Delta^2-(\epsilon'_{\mib k})^2}
\theta(\Delta-|\epsilon'_{\mib k}|)
-i{\rm sgn}(\pm\epsilon'_{\mib k})
\sqrt{(\epsilon'_{\mib k})^2-\Delta^2}
\theta(|\epsilon'_{\mib k}|-\Delta),
\end{equation}
where
$\epsilon'_{\mib k}=\epsilon+e{\mib v}_{\mib k}\cdot{\mib A}_0$
and 
$\theta(\cdot)$ is a step function.
The imaginary part of the
$\alpha\zeta^{'\pm}_{{\mib k},\epsilon}
\int_{FS}(1/\zeta^{'\pm}_{{\mib k},\epsilon})$ term in
Eq. (\ref{eq:approxzeta})
is finite for $|\epsilon|\ge \Delta-ev_FA_0$
in the clean case
($\alpha\ll \Delta$).
This is because
the term $e{\mib v}_{\mib k}\cdot{\mib A}_0$ is included 
in the integration
over the Fermi surface for the self-energy.
Then, ${\rm Re}[\text{ $\cdot$ }]$ in Eq. (\ref{eq:absbelow})
takes finite values only in the case of
$\epsilon+e{\mib v}_{\mib k}\cdot{\mib A}_0+\omega/2>
\Delta-ev_FA_0$ and
$\epsilon+e{\mib v}_{\mib k}\cdot{\mib A}_0-\omega/2<
-\Delta+ev_FA_0$.
This results in the finite absorption
for $\omega>2(\Delta-ev_FA_0)$ with a small but finite
impurity scattering ($\alpha\ne 0$). 
If we mistakenly adopt an approximation
$  \alpha\zeta^{'\pm}_{{\mib k},\epsilon}
\int_{FS}(1/\zeta^{'\pm}_{{\mib k},\epsilon})\simeq \alpha$
in Eq. (\ref{eq:approxzeta}) by neglecting
the self-energy effect, 
there is no absorption below $\omega<2\Delta$
at $T=0$ irrespective of the value of $\alpha>0$. 
This approximation results in the absorption edge
at $\omega=2\Delta$.

\section{Results of Numerical Calculations}

We numerically calculate the surface impedance
using Eqs. (\ref{eq:surfimp}) and (\ref{eq:currentq}).
We take $\Delta_0$ (the superconducting gap
in the case of $e{\mib v}_{\mib k}\cdot{\mib A}_0=0$
and $T=0$)
as the unit of energy ($\Delta_0=1$).
The dimensionless electron-phonon coupling constant $p$
in the gap equation is determined by Eq. (\ref{eq:gapeq})
under this condition.
The dependence of the superconducting gap $\Delta$
on $ev_FA_0$ is weak for $T/T_c\ll 1$
when $ev_FA_0<\Delta_0$. 
This is because
the shift of energy $|e{\mib v}_{\mib k}\cdot{\mib A}_0|$
is small as compared to the range of integration
over $\epsilon$ in Eq. (\ref{eq:gapeq})
($ev_kA_0\ll\omega_D$; $\omega_D$ is
the Debye frequency and we put $\omega_D=8\Delta_0$).
The calculations are performed at $T=0$ because
the contribution from the thermally excited quasiparticles
to the conductivity is small
and $\Delta\simeq \Delta_0$
at low temperatures.
  The results given below
  are obtained by the calculation with ${\mib A}_0$
  being parallel to $\tilde{E}^x_{\omega}(q)$
  (the perpendicular case
  corresponds to the term without vertex corrections
  as mentioned in Sect. 2.1).

When the values of $2\Delta_0/k_BT_c=3.54$ and
$T_c=1.1$ K are used,
the scale of length is numerically given by
$1$ $\mu\text{m}\simeq 0.85\times 10^{-3}$
as we put $\hbar=c=1$.
However, this scale does not appear explicitly
in the numerical calculation, and  
only dimensionless quantities, $\xi_0q$ and
$\xi_0/\lambda_0$, are needed in calculations
($\xi_0=v_F/\pi\Delta_0$ is the coherence length).
This is because Eq. (\ref{eq:surfimp}) is
rewritten as
  \begin{equation}
    Z_S=
    4\pi i\omega\int\frac{d(\xi_0q)}{2\pi}
    \frac{-2\xi_0}
         {(\xi_0q)^2
           -4\pi i\xi_0^2\tilde{j}_{\omega}^x(q)/\tilde{E}^x_{\omega}(q)},
\label{eq:surfimpxi}
  \end{equation}
and  $\tilde{j}^x_{\omega}(q)$ multiplied by $\xi_0^2$
  is proportional to $(\xi_0/\lambda_0)^2$.
  [If we take into account a displacement current term in the Maxwell equation
  Eq. (\ref{eq:zmaxwelleq}), $(\xi_0q)^2$ is replaced by
  $(\xi_0q)^2-(\xi_0\omega)^2$.
  When $\omega\sim 2\Delta_0$ and $\xi_0\sim 1$ $\mu\text{m}$,
  $(\xi_0\omega)^2\sim 3.4\times 10^{-6}$ and then
  this term can be omitted.]

  \subsection{Surface resistance and one-particle
    spectrum}
  
\begin{figure}
  \includegraphics[width=11.5cm]{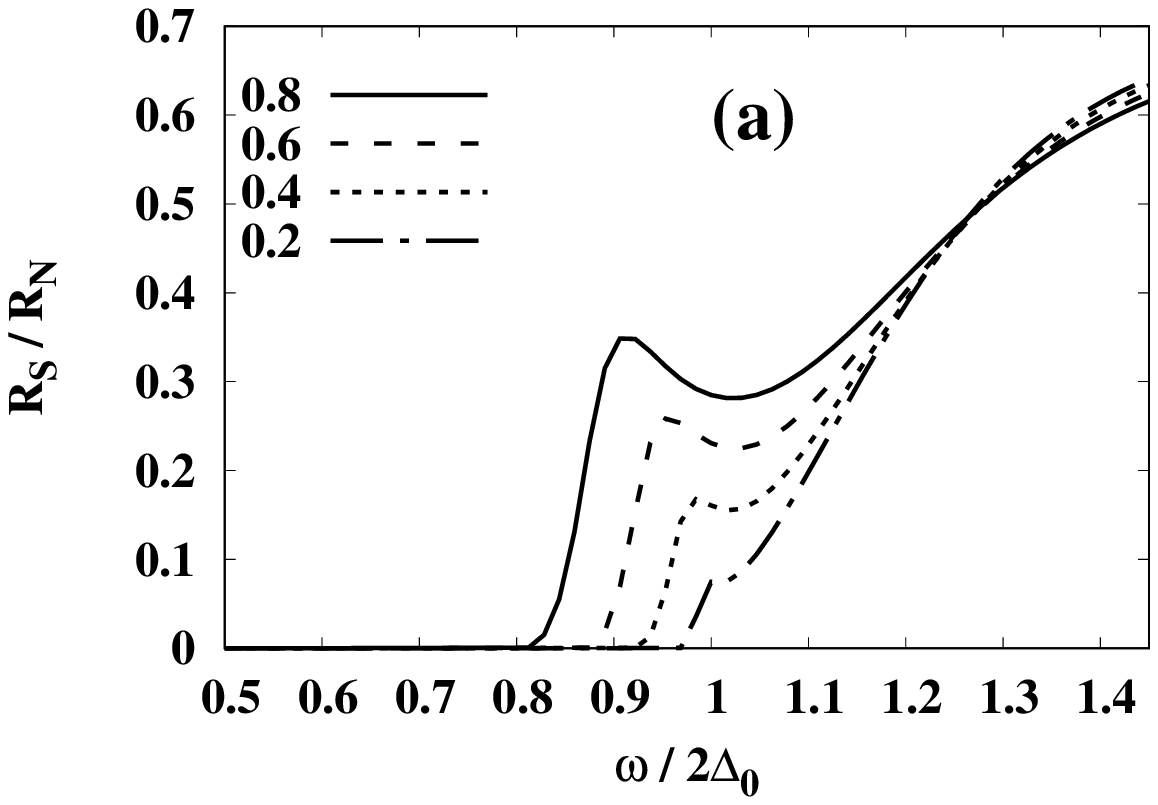}
  \includegraphics[width=11.5cm]{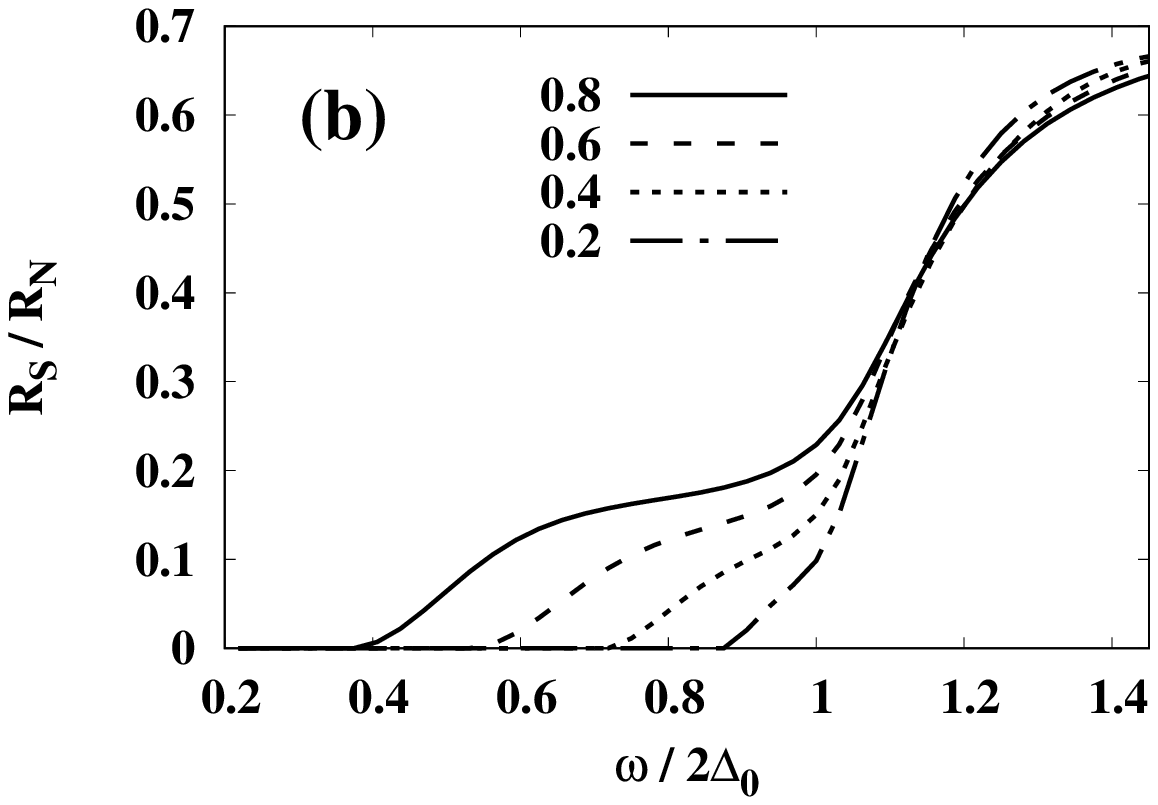}
  \caption{
    \label{fig:1}
    Dependence of
    surface resistance on $\omega$
    for various values of $ev_FA_0$
    ($=0.2$, $0.4$, $0.6$, and $0.8$).
$\xi_0/\lambda_0=4.0$.  
 (a)   $\alpha=5.0$ and (b) $\alpha=0.3$.
  }
\end{figure}

  The calculated surface resistance
($R_S={\rm Re}Z_S$)
  divided by the quantity in the normal state
  ($R_N={\rm Re}Z_S|_{\Delta=0}$)
  is shown Fig.~\ref{fig:1}.
The surface resistance takes finite values
below $2\Delta_0$ ($\omega<2\Delta_0$),
and it has a peak in the dirty case ($\alpha=5.0$).
In the relatively clean case ($\alpha=0.3$),
there is a broad tail in $R_S$ for $\omega<2\Delta_0$.
This difference in the spectrum at the gap edge
between the dirty and clean cases is 
related to that in the one-particle spectrum.

\begin{figure}
  \includegraphics[width=11.5cm]{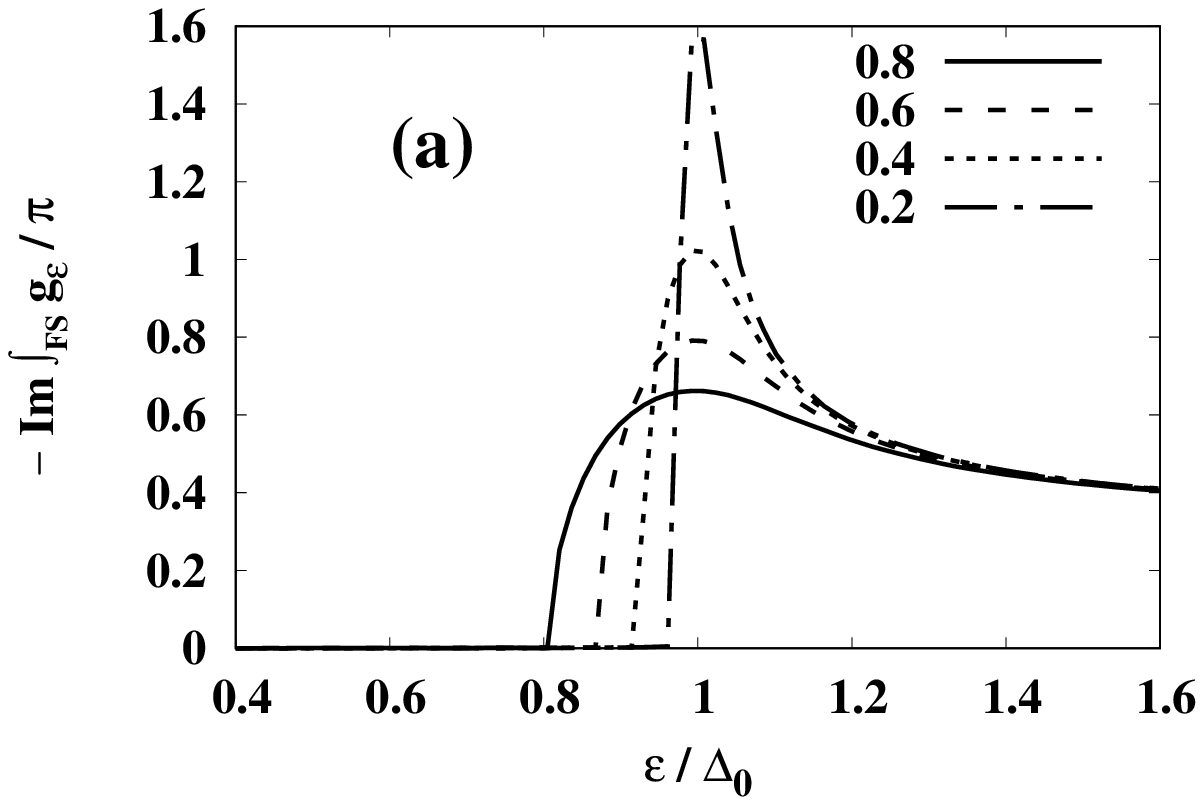}
  \includegraphics[width=11.5cm]{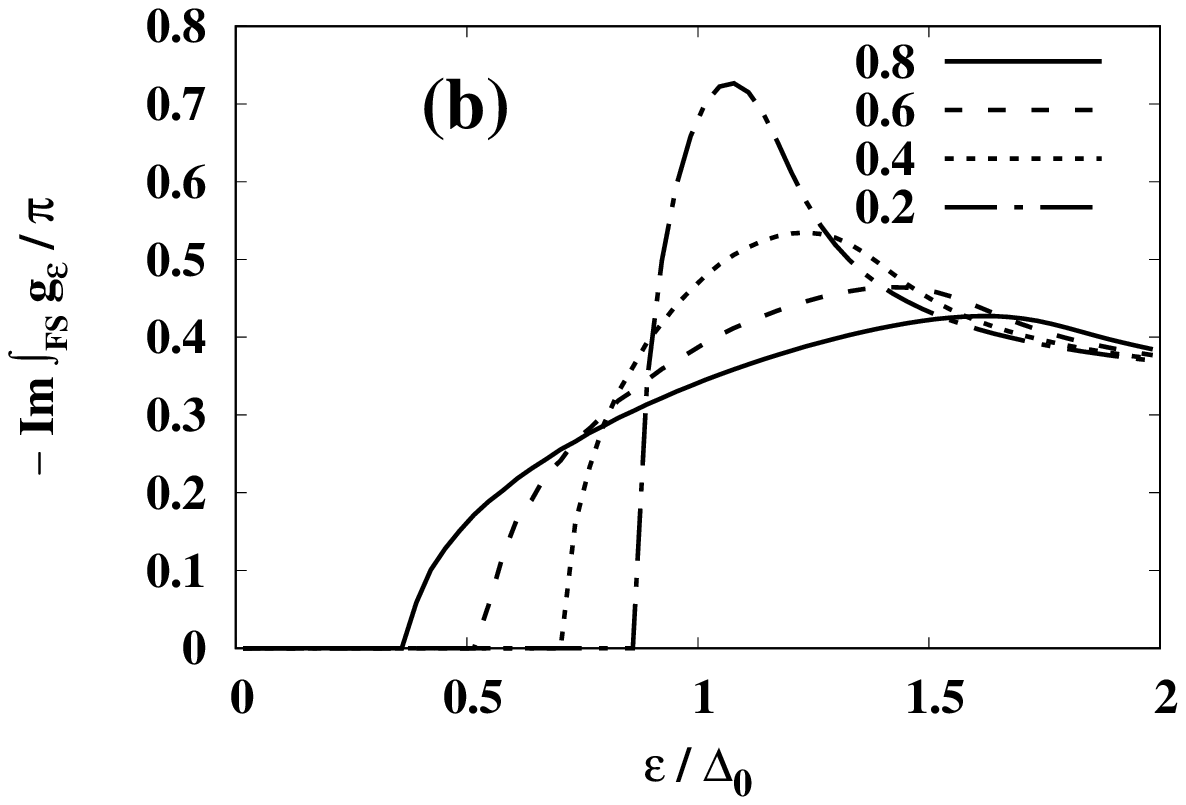}
  \caption{
    \label{fig:2}
    Density of states for $ev_FA_0=0.2$, $0.4$, $0.6$, and $0.8$.
  (a)  $\alpha=5.0$ and (b) $\alpha=0.3$.
  }
\end{figure}

The calculated results of
the density of states
($-{\rm Im}\int_{FS}g^{+}_{{\mib k},\epsilon}/\pi$)
are shown in Fig.~\ref{fig:2}.
In the relatively clean case,
the self-energy correction to the one-particle spectrum
is small, and 
the renormalization of the density of states is small.
Then, the shift of the spectrum is determined
by $e{\mib v}_{\mib k}\cdot{\mib A}_0$ and 
the gap edge is close to $\Delta-ev_FA_0$,
which is the gap edge
in the case with no impurity scattering ($\alpha=0$).
On the other hand,
in the dirty case,
the impurity scattering renormalizes 
the shift of the spectrum, and 
the gap edge shifts to a higher energy,
$|\epsilon|\sim E_g=\Delta[1-(2\alpha_p/\Delta)^{2/3}]^{3/2}$,
as in the case with paramagnetic impurities.~\cite{skalski}
[$\alpha_p$ is the pair breaking parameter~\cite{maki}: 
$\alpha_p=(ev_FA_0)^2/6\alpha$.  
  In the case of $ev_FA_0=0.8$ and $\Delta=1$,
    $E_g\simeq 0.83$ and $0.16$ 
($\alpha_p\simeq 0.02$ and $0.35$) for $\alpha=5.0$
and $0.3$, respectively.
In the clean case ($\alpha< \Delta_0$),
this expression is not valid
and the gap edge is simply related to
the shift ($ev_FA_0$) as noted above.]

\begin{figure}
  \includegraphics[width=11.5cm]{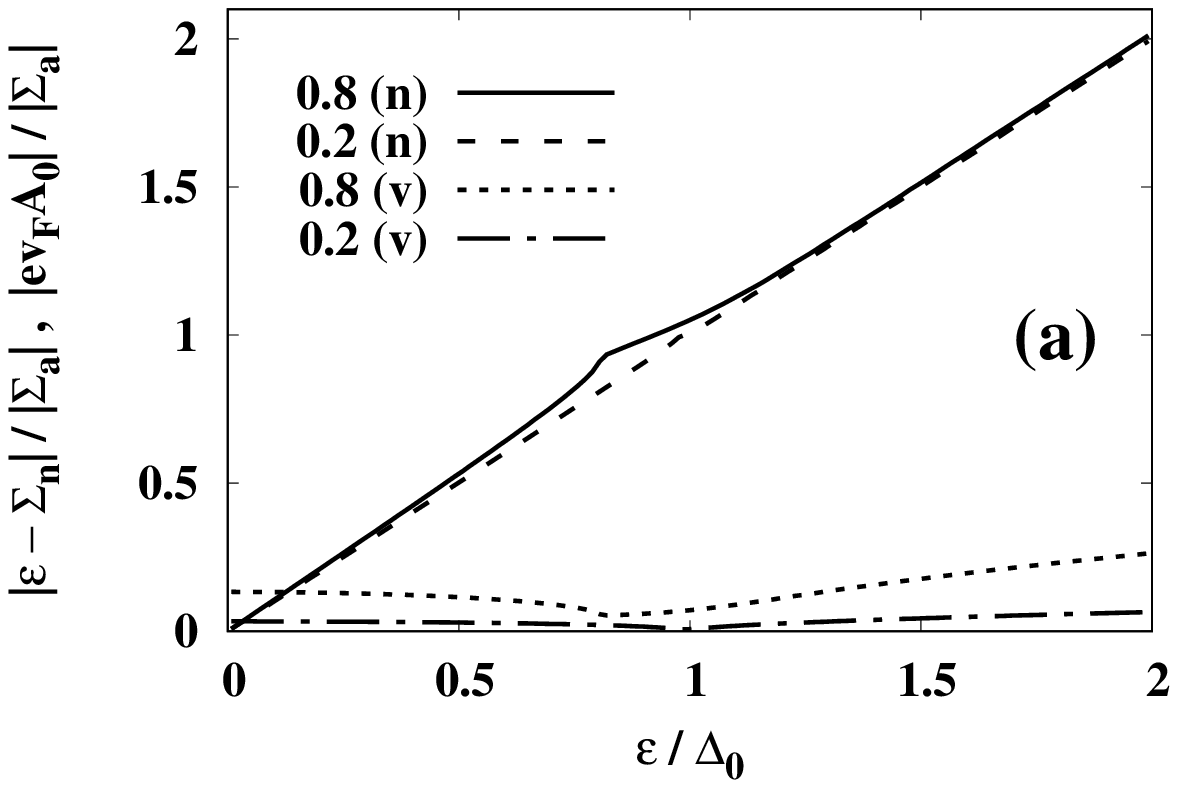}
  \includegraphics[width=11.5cm]{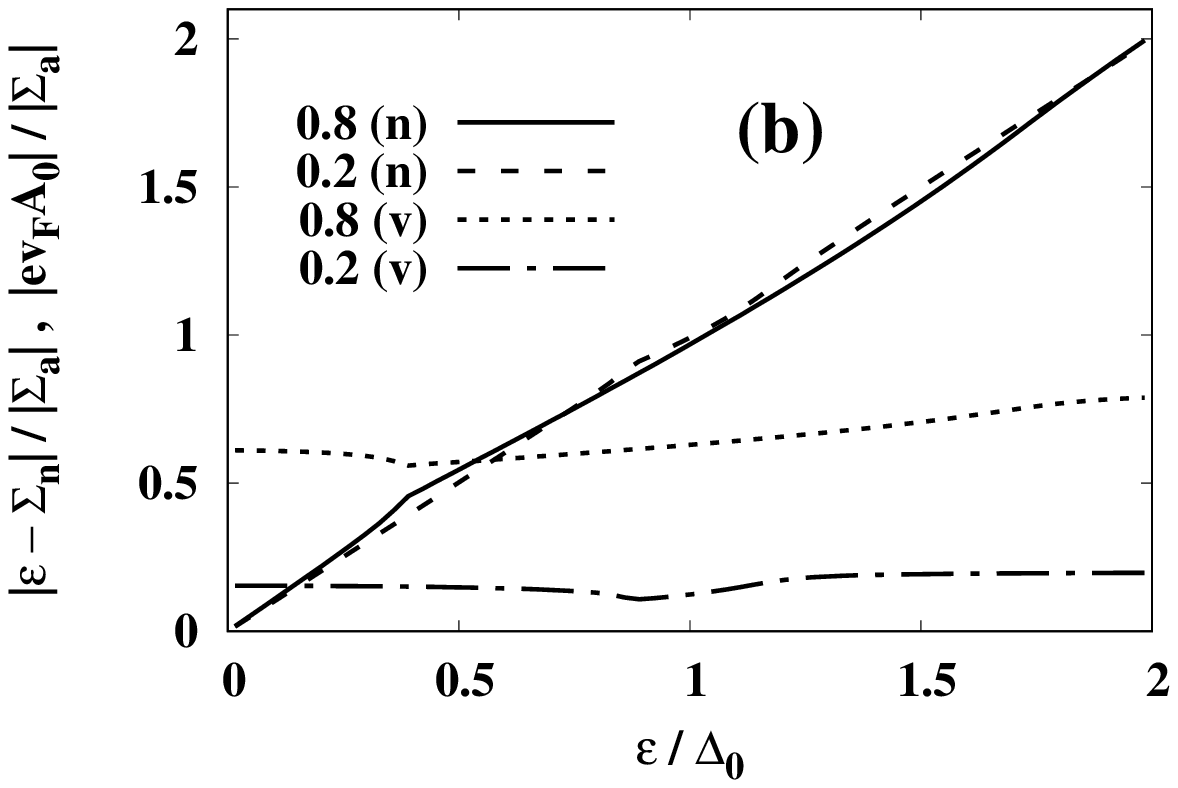}
  \caption{
    \label{fig:3}
    Dependences of 
    $|\epsilon-\Sigma_n(\epsilon)|/
    |\Sigma_a(\epsilon)|=(\text{n})$ 
    and 
        $|ev_FA_0|/
    |\Sigma_a(\epsilon)|=(\text{v})$ 
    on $\epsilon$
    for $ev_FA_0=0.2$ and $0.8$.
   (a) $\alpha=5.0$ and (b) $\alpha=0.3$.
  }
\end{figure}

The effects mentioned above can be understood 
by calculating quantities, which include the self-energy
corrections, such as 
$|\epsilon-\Sigma_n(\epsilon)|/|\Sigma_a(\epsilon)|$
and $|ev_FA_0|/|\Sigma_a(\epsilon)|$
[$\Sigma^{\pm}_a(\epsilon)=\Delta-\alpha\int_{FS}f^{\pm}_{{\mib k},\epsilon}$
  given by Eq. (\ref{eq:equiliselfenergy})]
shown in Fig.~\ref{fig:3}.
Equation (\ref{eq:onepartg}) indicates that
the quantities shown in Fig. 3 determine
the behavior of the density of states.
In the case of $ev_FA_0=0$,
$|\epsilon-\Sigma_n(\epsilon)|/
|\Sigma_a(\epsilon)|=
|\epsilon|/\Delta$ 
irrespective of the values of $\alpha$
(Anderson's theorem).~\cite{anderson59}
Figure 3 shows that this relation approximately
holds for finite values of
$ev_FA_0\ne 0$. 
On the other hand,
the values of $|ev_FA_0|/
|\Sigma_a(\epsilon)|$
depend on $\alpha$ and become small for $\alpha>1$.
The renormalization of the density of states results from this
dependence of $|ev_FA_0|/
|\Sigma_a(\epsilon)|$ on $\alpha$.
(The shift of the gap edge in the density of states
has been shown in previous studies.~\cite{sanchez-canizares})

\subsection{Vertex correction and 
  spatial variation of conductivity}

\begin{figure}
  \includegraphics[width=9.5cm]{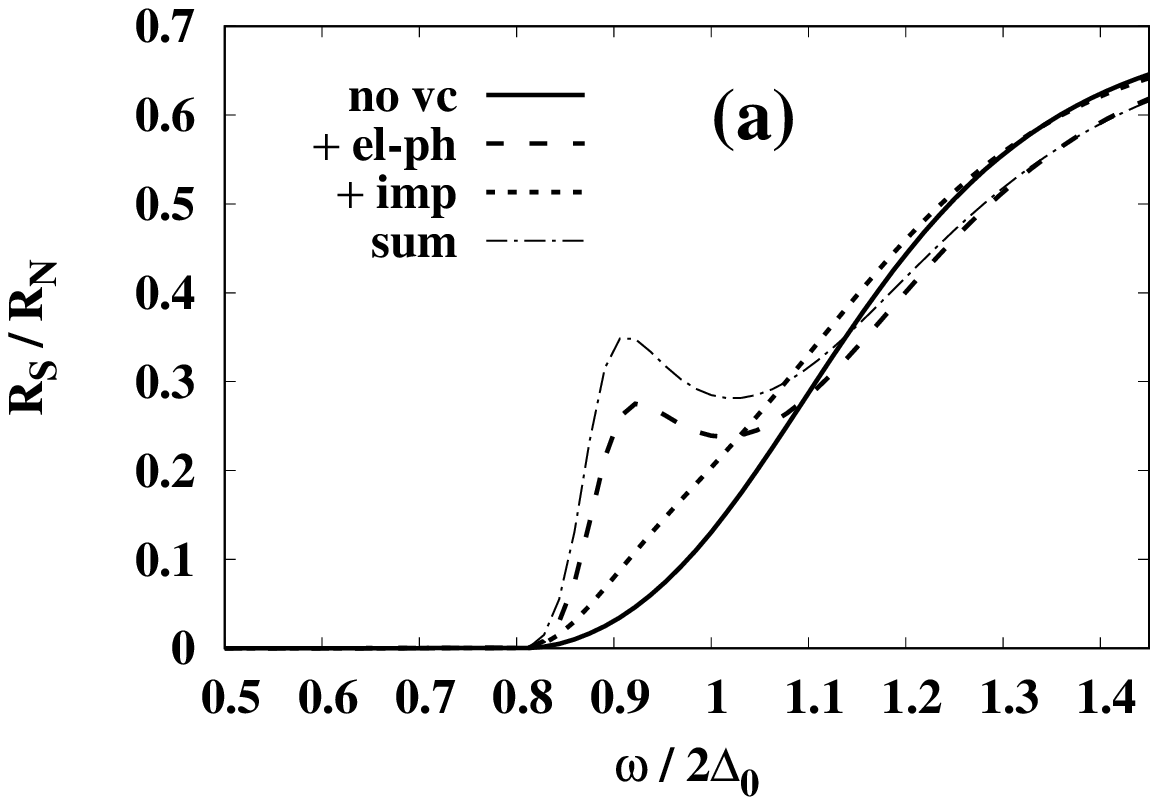}
  \includegraphics[width=9.5cm]{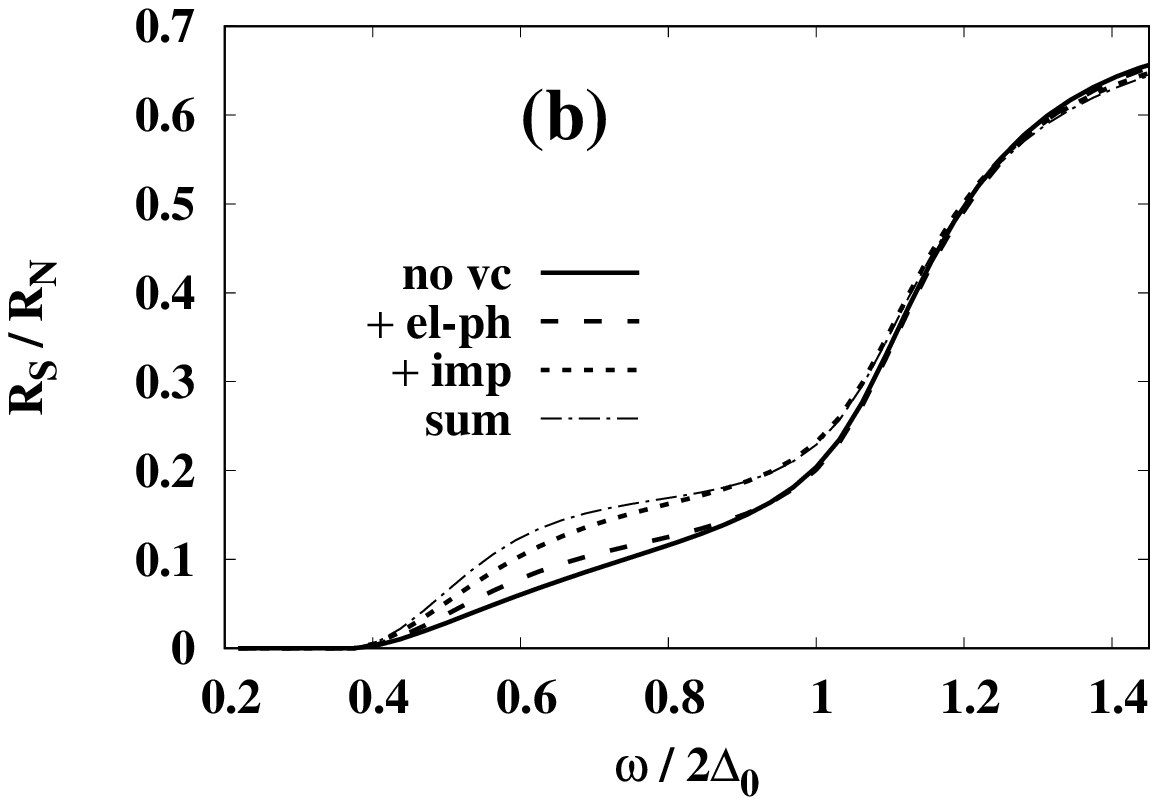}
  \includegraphics[width=9.5cm]{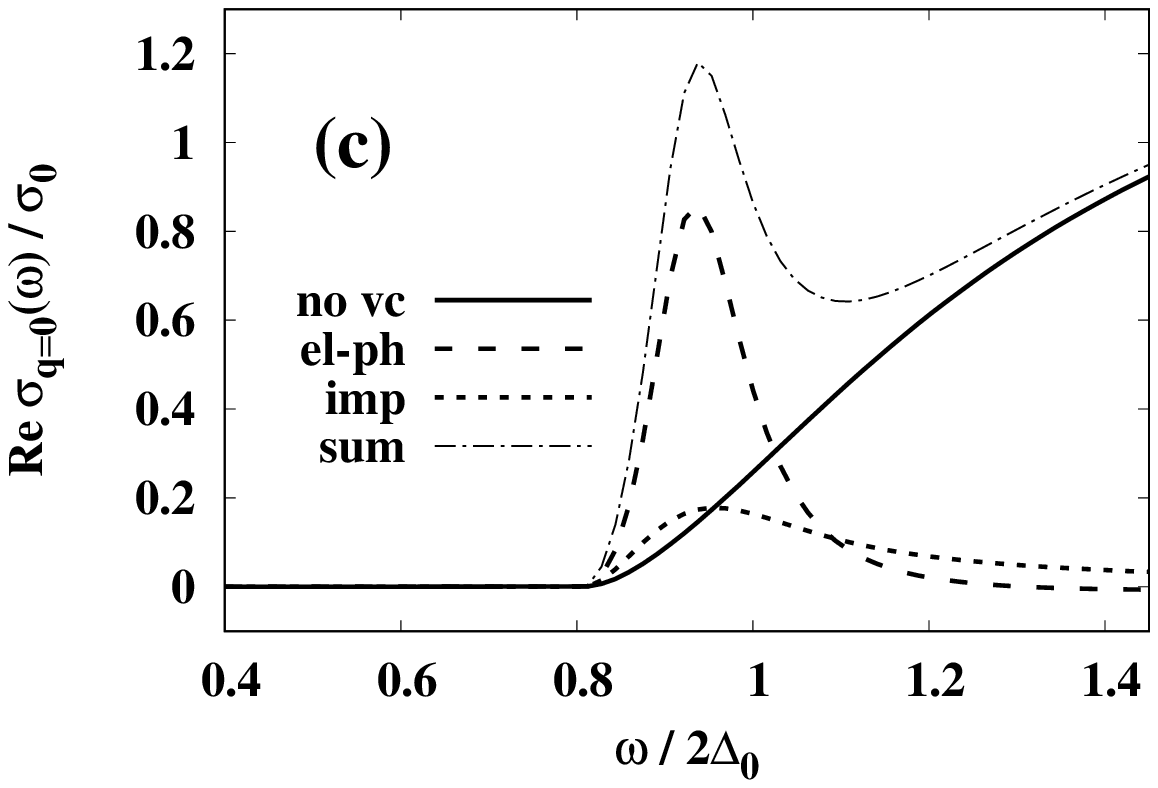}
  \includegraphics[width=9.5cm]{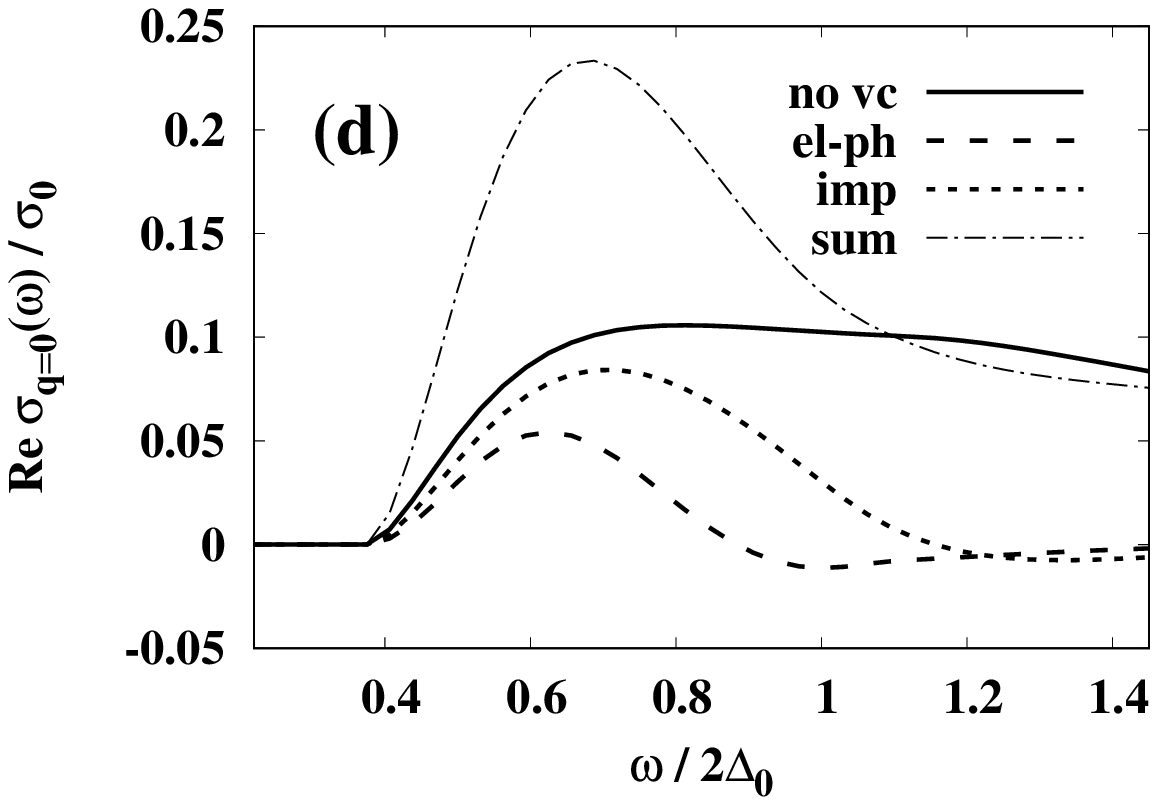}
  \caption{
    \label{fig:4}
In (a) and (b),
    `no vc', `+el-ph', `+imp', and `sum' mean 
    the calculated results 
    with the current
    $\tilde{j}_{\omega}^x(q)$ in 
    Eq. (\ref{eq:surfimpxi})
    replaced by 
    $\tilde{j}_{\omega}^{x(0)}(q)$,
    $\tilde{j}_{\omega}^{x(0)}(q)+\tilde{j}_{\omega}^{x(ep)}(q)$,
    $\tilde{j}_{\omega}^{x(0)}(q)+\tilde{j}_{\omega}^{x(im)}(q)$,
    and $\tilde{j}_{\omega}^x(q)$
    (unchanged), respectively.
In (c) and (d),
    `no vc', `el-ph', `imp', and `sum' mean 
    the calculated results of
    ${\rm Re}[\tilde{j}^{(0)x}_{\omega}(q)/\tilde{E}^x_{\omega}(q)]_{q=0}
    /\sigma_0$,
    ${\rm Re}[\tilde{j}_{\omega}^{x(ep)}(q)/\tilde{E}^x_{\omega}(q)]_{q=0}
    /\sigma_0$,
    ${\rm Re}[\tilde{j}_{\omega}^{x(im)}(q)/\tilde{E}^x_{\omega}(q)]_{q=0}
    /\sigma_0$,
    and 
    ${\rm Re}[\tilde{j}^x_{\omega}(q)/\tilde{E}^x_{\omega}(q)]_{q=0}
    /\sigma_0$, respectively.
 $ev_FA_0=0.8$ and $\xi_0/\lambda_0=4.0$.
    $\alpha=5.0$ for (a) and (c). $\alpha=0.3$ for (b) and (d).
      }
\end{figure}

The physical origin of the peak structure shown in
Fig. 1 can be understood by calculating the surface resistance
and the conductivity with the use of the decomposition
in Eq. (\ref{eq:currentq}).
The calculated results of several terms
for $R_S/R_N$
and ${\rm Re}\sigma_q(\omega)/\sigma_0$
($\sigma_0=e^2 n_e \tau/m$ with $\tau=1/2\alpha$) 
are shown in 
Fig.~\ref{fig:4}.
These results show that the peak of the surface resistance
in the dirty case ($\alpha=5.0$)
originates from the vertex correction
by the electron--phonon interaction (amplitude fluctuation).
On the other hand, in the clean case ($\alpha=0.3$),
the vertex correction terms
are small as compared to the term with no vertex correction,
and the contribution of the amplitude mode to
the conductivity is relatively small.

\begin{figure}
  \includegraphics[width=11.5cm]{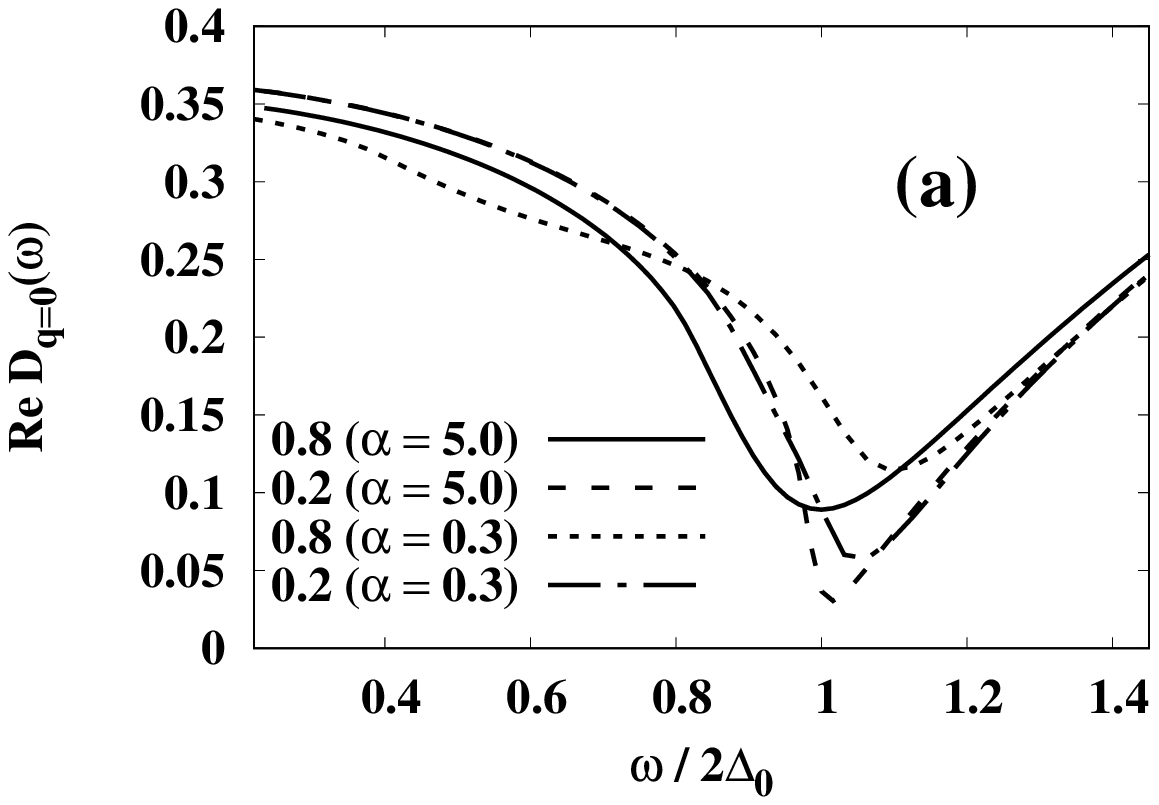}
  \includegraphics[width=11.5cm]{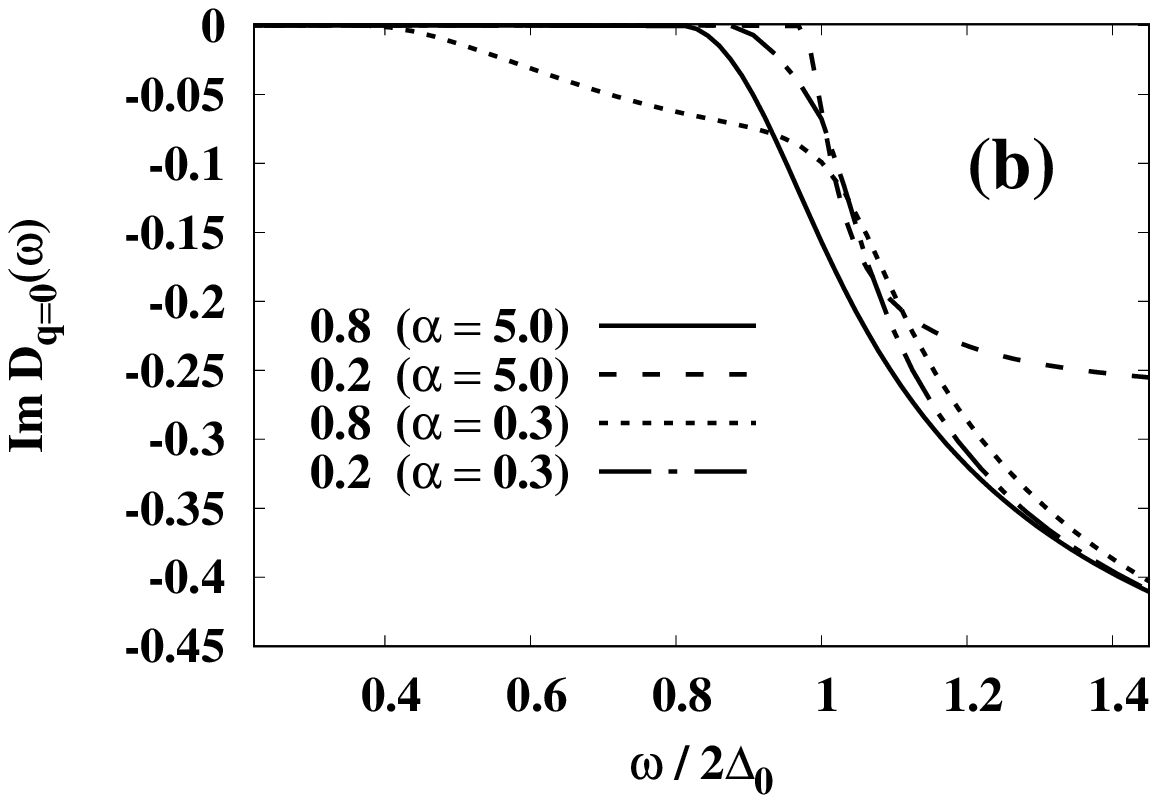}
  \caption{
    \label{fig:5}
    Dependences of
    (a) real and (b) imaginary parts of
    $D_{q=0}(\omega)$ on $\omega$ for
    $ev_FA_0=0.2$ and $0.8$.
    $\alpha=0.3$ and $5.0$.
      }
\end{figure}

The quantity representing the amplitude mode
is the denominator of Eq. (\ref{eq:ampelphvc}):
\begin{equation}
 D_{q}(\omega):=           1+p\int\frac{d\epsilon}{2\pi i}
    \left[T^h_{\epsilon^-}
\kappa^{++(v0)}_{\epsilon^+,\epsilon^-}(q)
    -T^h_{\epsilon^+}
    \kappa^{--(v0)}_{\epsilon^+,\epsilon^-}(q)
    +(T^h_{\epsilon^+}-T^h_{\epsilon^-})
    \kappa^{+-(v0)}_{\epsilon^+,\epsilon^-}(q)
    \right].
    \label{eq:ampmode}
\end{equation}
The real and imaginary parts of $D_{q=0}(\omega)$
are shown in Fig.~\ref{fig:5}.
The results for 
the real part
${\rm Re}D_{q=0}(\omega)$ show that
its minimum value around $\omega\simeq 2\Delta$
for the dirty case ($\alpha=5.0$)
is smaller than that for the clean case ($\alpha=0.3$).
The damping 
${\rm Im}D_{q=0}(\omega)$ shows a different
broadness between the clean and dirty cases, which
originates from the one-particle density of states
as shown in Fig. 2.
This behavior of the amplitude mode partly leads to
a large (small) electron--phonon vertex correction in
the dirty (clean) case as shown in Fig. 4.
(This behavior is consistent with
the result for the superconductors with paramagnetic impurities
shown in Fig. 3 of Ref. 37
because the large (small) values of $ev_FA_0$ ($\alpha$) 
correspond to the large values of $\alpha_p$ in 
Ref. 37.)

\begin{figure}
  \includegraphics[width=11.5cm]{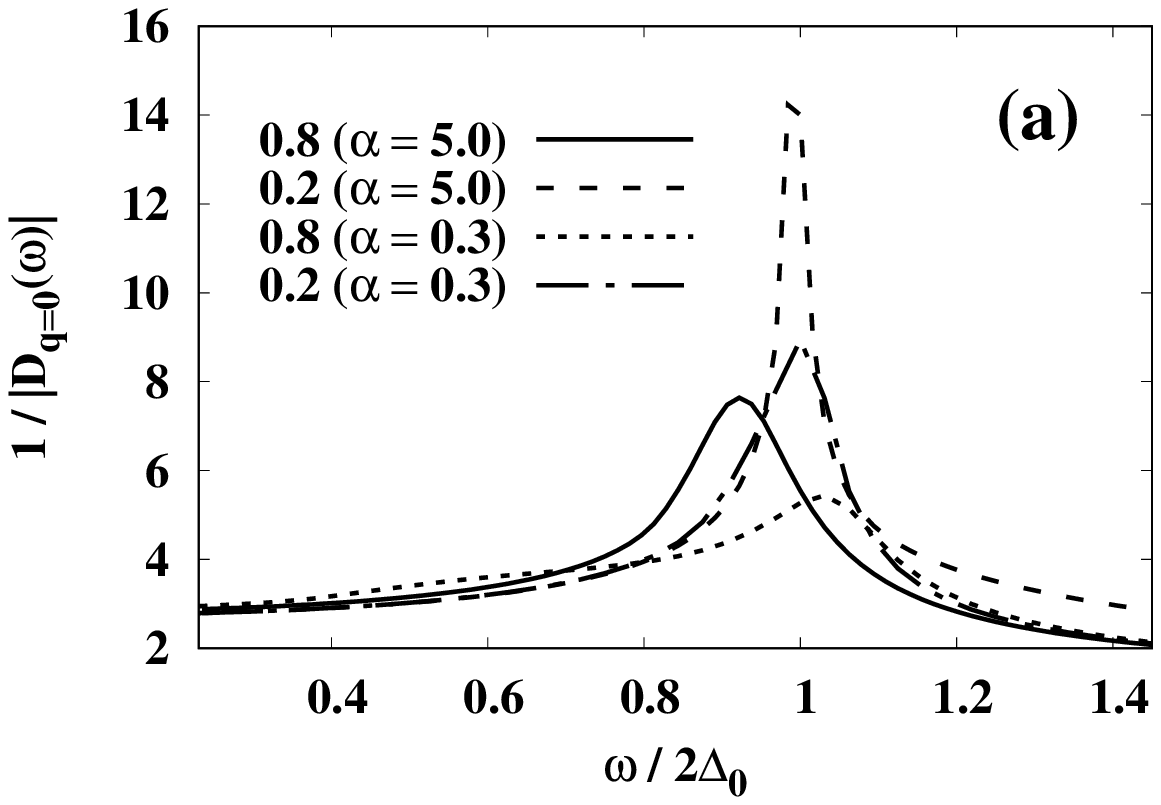}
  \includegraphics[width=11.5cm]{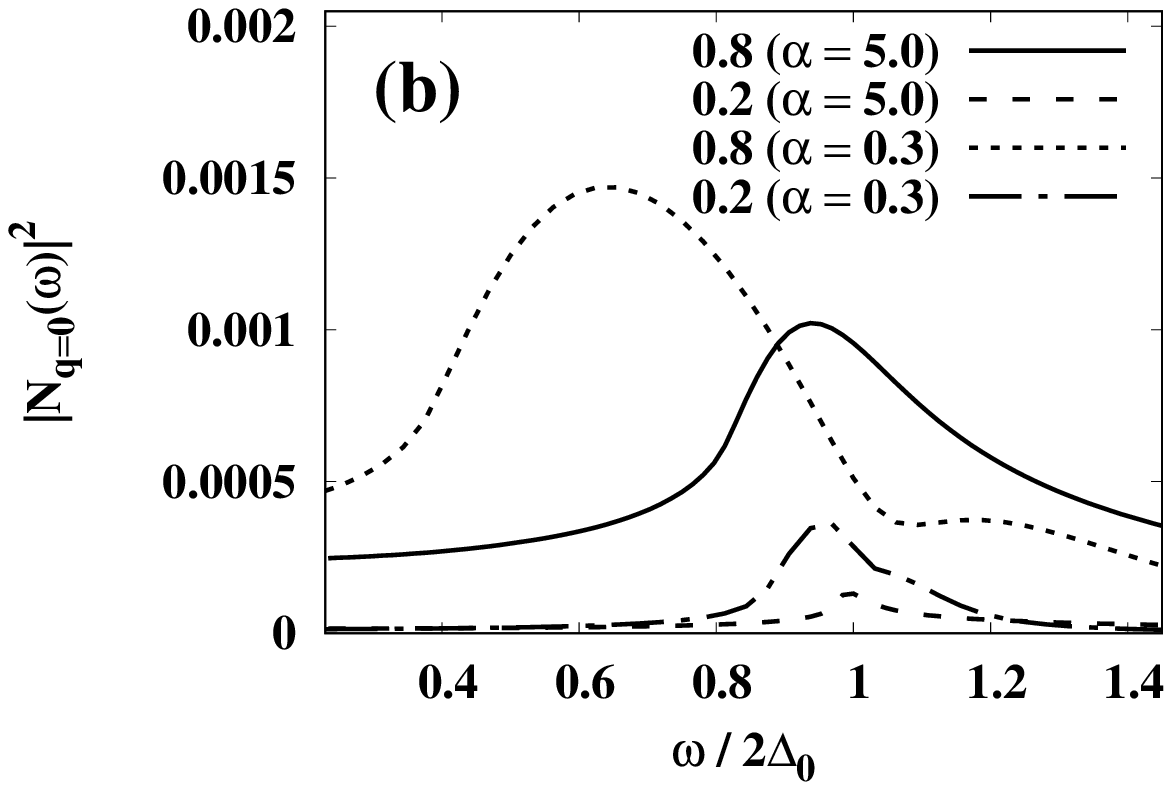}
  \caption{
    \label{fig:6}
    Dependences of
    (a)  $1/|D_{q=0}(\omega)|$ 
    and (b) $|N_{q=0}(\omega)|^2$
    on $\omega$ for
    $ev_FA_0=0.2$ and $0.8$.
    $\alpha=0.3$ and $5.0$.
      }
\end{figure}

From Eqs. (\ref{eq:currentkappa}), (\ref{eq:sigelph}), (\ref{eq:ampelphvc}),
and (\ref{eq:ampmode}), 
the current including the amplitude mode is rewritten as
\begin{equation}
  \tilde{j}^{x(ep)}_{\omega}(q)
  =\frac{3p [N_q(\omega)]^2}{2\lambda_0^2 D_q(\omega)}
  \tilde{A}^x_{\omega}(q)
\end{equation}
with
\begin{equation}
  N_q(\omega):=
  \int\frac{d\epsilon}{4\pi i}
    \left[T^h_{\epsilon^-}
    \kappa^{++(v1)}_{\epsilon^+,\epsilon^-}(q)
    -T^h_{\epsilon^+}
    \kappa^{--(v1)}_{\epsilon^+,\epsilon^-}(q)
    +(T^h_{\epsilon^+}-T^h_{\epsilon^-})
    \kappa^{+-(v1)}_{\epsilon^+,\epsilon^-}(q)
\right].
\end{equation}
The dependences of 
$1/|D_{q=0}(\omega)|$ and $|N_{q=0}(\omega)|^2$
on frequency are shown in
Fig.~\ref{fig:6}.
There is a peak of $1/|D_{q=0}(\omega)|$
around $\omega\simeq 2\Delta$ in the dirty and clean cases.
The peak values in the dirty case are larger
than those in the clean case as expected from the
results shown in Fig. 5.
The term in the numerator of $\tilde{j}^{x(ep)}_{\omega}(q)$,
$N_{q}(\omega)$, has a large weight
around the frequency that corresponds to
the gap edge in the one-particle spectrum as in Fig. 2.
[This is because the coupling to the external field
$ev_{\mib k}^x\tilde{A}^x_{\omega}(q)$
  is effective in the direction of momentum ${\mib k}$
  in which there is a large energy shift 
$|ev_{\mib k}^x A^x_0|$ ($|v_{\mib k}^x|\simeq v_F$)
in the energy spectrum.
The dip around $\omega\simeq 2\Delta$ corresponds to
the ineffective coupling to the external field
and a small energy shift.
On the other hand, in the dirty case, these shifts are averaged out
in the energy spectrum owing to the impurity scattering.]
Then, in the clean case,
there exists a discrepancy between
the frequencies at which $1/D_q(\omega)$
and $N_q(\omega)$ take their largest values.
This is the reason why
the electron--phonon vertex correction
is not effective for the surface resistance
in the clean case 
as shown in
Fig. 4.

\begin{figure}
  \includegraphics[width=11.5cm]{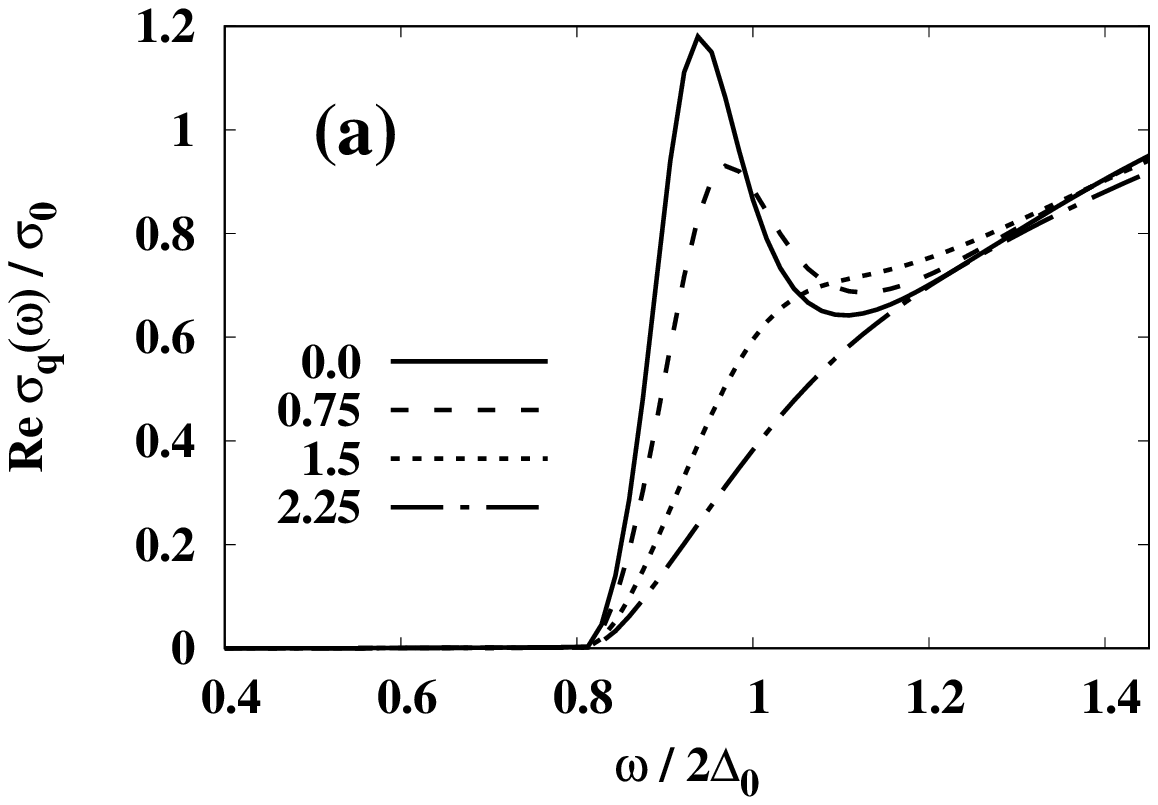}
\includegraphics[width=11.5cm]{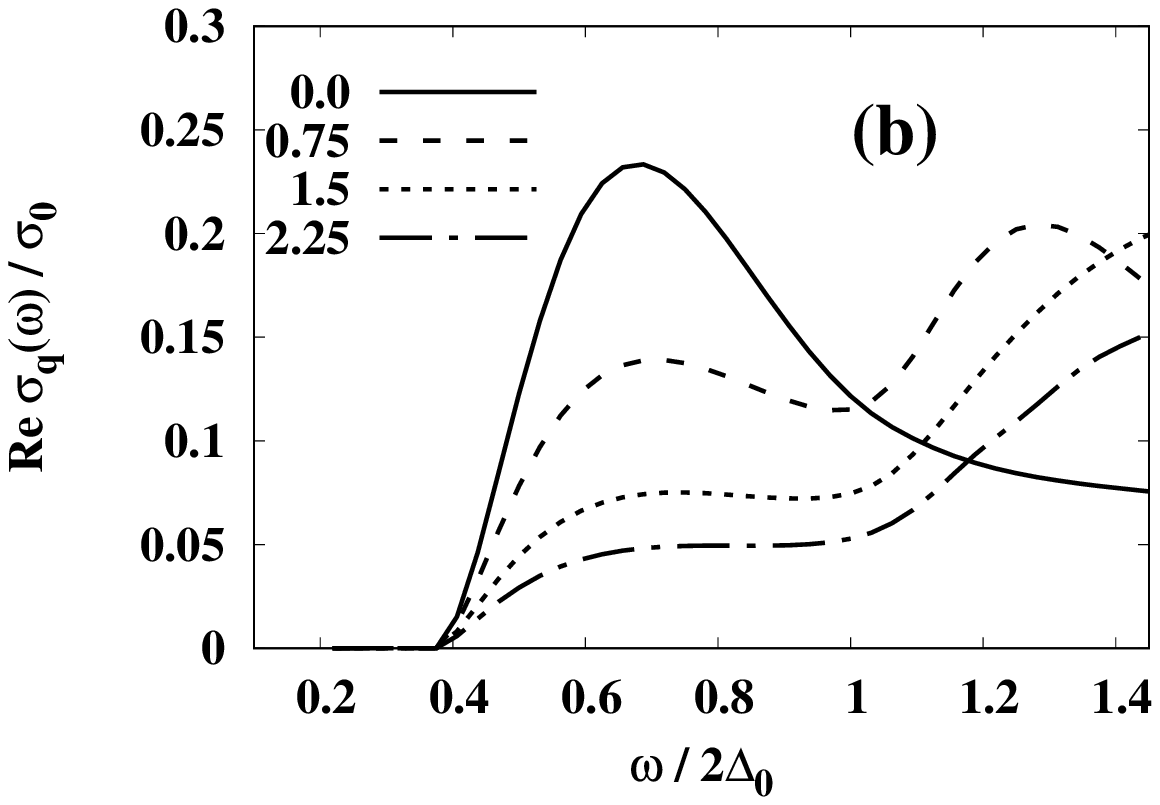}
  \caption{
    \label{fig:7}
    Dependences of absorptive part of conductivity
    ${\rm Re}\sigma_{\mib q}(\omega)/\sigma_0$
    on $\omega$ 
    for various values $\xi_0q$ ($=0.0$, $0.75$, $1.5$, and $2..25$).
        $ev_FA_0=0.8$. 
    (a) $\alpha=5.0$ and (b) $\alpha=0.3$
  }
\end{figure}

The above calculations of $R_S$ are performed in the nonlocal case
($\xi_0/\lambda_0>1$), and in this case,
the spatial variation
(a dependence on $q$)
of conductivity is expected to be important in the calculation of 
surface impedance.
The real part of conductivity for several 
$\xi_0q$ values is shown in 
Fig.~\ref{fig:7}.
Here, $\sigma_q(\omega)=\tilde{j}_{\omega}^x(q)
/\tilde{E}_{\omega}^x(q)$. 
${\rm Re}\sigma_{\mib q}(\omega)/\sigma_0$
does not depend on 
$\xi_0/\lambda_0$ because a factor $1/\lambda_0^2$
in the coefficient of $\tilde{j}_{\omega}^x(q)$
cancels with that in $\sigma_0\propto 1/\lambda_0^2$.
There is a peak in the absorption spectrum for 
$q=0$ 
even in the relatively clean case ($\alpha=0.3$).
In the nonlocal case ($\xi_0/\lambda_0>1$),
however,
the surface impedance
includes the conductivity over 
a wide range of
wave numbers ($\xi_0q\sim \xi_0/\lambda_0>1$)
in the integral [Eq. (\ref{eq:surfimpxi})]. 
Then, the broad peak in the conductivity
is smeared out in the surface resistance
in the clean case.
On the other hand, in the dirty case,
there is a sharp peak in the conductivity 
because of the amplitude mode and
the narrow spectrum in the density of states
as in Fig. 2.
This results in the peak structure in the surface resistance
as shown in Fig. 1.

\begin{figure}
  \includegraphics[width=11.5cm]{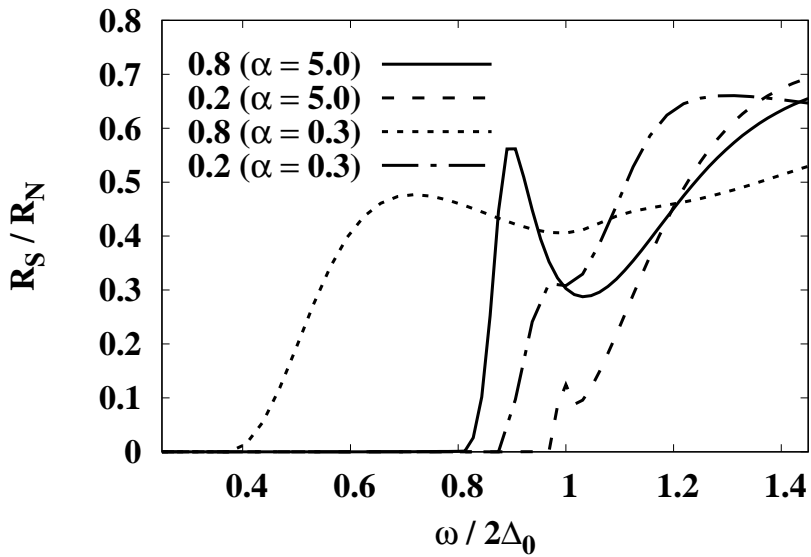}
  \caption{
    \label{fig:8}
Surface resistance in local case ($\xi_0/\lambda_0=0.25$)
with $ev_FA_0=0.8$ and $ev_FA_0=0.2$
for $\alpha=5.0$ and $\alpha=0.3$. 
  }
\end{figure}

If we consider the local case
($\xi_0/\lambda_0<1$),
the conductivity at small $q$
($\xi_0q\sim \xi_0/\lambda_0<1$) mainly contributes to $Z_S$
in the integration of Eq. (\ref{eq:surfimpxi}). 
Then, the peak structure in the surface resistance can be seen even in
the clean case.
The calculated surface resistance for 
$\xi_0/\lambda_0=0.25$
is shown in Fig.~\ref{fig:8}.
As compared with Fig. 1, the peak structure becomes sharp in the dirty
case ($\alpha=5.0$), and a broad peak can be seen in
the clean case ($\alpha=0.3$).
The physical origin of these peaks is different between
the dirty and clean cases as shown in Fig. 4.
The amplitude mode can be seen in the dirty case, but
it is relatively small as compared to other terms
in the clean case.
The broad peak in the clean case originates from
an interlocking between the coupling to
the external field and the energy shift as
discussed above for $N_q(\omega)$ (Fig. 5).

\section{Summary and Discussion}

We calculated the surface resistance
under a uniform and static external field.
Nonlocality is taken into account 
for the dirty and clean superconductors.
Although previous studies have concentrated on
the local and dirty case,
we showed that the peak caused by the amplitude mode
appears even in the nonlocal case.
The ineffectiveness of the amplitude mode
in the clean case is also clarified,
and the reason for this is that
the weight of the effective coupling to the external field 
shifts away from that of the amplitude mode.

The calculated results are consistent 
with the experiment~\cite{budzinski,budzinskiB} qualitatively
in both dirty and clean cases.
  (As mentioned below, if the effect of impurity scattering
  is extremely small, it is possible that other scattering processes, 
  such as electron--phonon interactions, are quantitatively
  effective for the absorption spectrum.)
With use of $\xi_0$
and the mean free path ($l=v_F\tau$),
$\alpha=(\pi\Delta_0/2)(\xi_0/l)$, and 
the values of $\xi_0/l\sim 7$ and $0.04$
for silver-doped Al and pure Al~\cite{budzinski,budzinskiB}
correspond to 
$\alpha\simeq 11.0$ and $0.063$, respectively.
$\xi_0/\lambda_0>\sqrt{2}$ in both cases.
We assumed the static external field
to be uniform.
If the external field exists in $z<L$,
the relation of $ev_FA_0$ to the magnetic field $H$
is written as
$ev_FA_0/\Delta_0=(\pi L/2\sqrt{2}\lambda_0)(H/H_c)$
($H_c=\hbar c/2\sqrt{2}e\xi_0\lambda_0$). 
Then, $ev_FA_0/\Delta_0=0.8$ corresponds to $H/H_{c}\simeq 0.72$
in the case of $L=\lambda_0$.
This estimation indicates that
the uniform $A_0$ is a
first step as an approximation, and 
the problem with nonuniform $A_0$ $(\propto e^{-|z|/\lambda_0})$ is a future
issue to be solved by calculating the differential equation
numerically.

Another issue to be considered
is
that the dependence of the frequency of the gap edge
on the directions of external fields is strong
for the clean case
in the surface resistance.~\cite{budzinski,budzinskiB}
In our calculation,
there is no difference in the frequency of the gap edge
between the surface resistances for $E_x^{\omega}\parallel {\mib A}_0$
and $E_x^{\omega}\perp {\mib A}_0$, which correspond to
the term `sum' and `no vc' in Fig. 4(b), respectively.
This is because 
the effect of 
impurity scattering is integrated over the
Fermi surface and the self-energy does not
depend on the direction in the momentum space.
A possible reason for this discrepancy is
the smallness of $\alpha\simeq 0.063$ in experiments.
It is possible that 
the interaction between electrons and acoustic phonons
predominates over the impurity scattering effect
in the extreme clean case.
In this case,
the damping rate is written as 
${\rm Im}\hat{\Sigma}_{{\mib k},\epsilon}^{+(ap)}
=(p'/4\omega_D^2)\int d\epsilon' (\epsilon-\epsilon')
|\epsilon-\epsilon'|(T^h_{\epsilon'}+1/T^h_{\epsilon-\epsilon'})
\hat{\tau}_3    {\rm Im}\hat{g}^+_{{\mib k},\epsilon'}
\hat{\tau}_3$~\cite{kopnin}
($p'$ is the contribution to $p$ by acoustic phonons
and $p'< 0.56$).
This quantity depends on the direction of ${\mib A}_0$
through $e{\mib v}_{\mib k}\cdot{\mib A}_0$
and makes the surface resistance dependent
on the direction of the external field.

Our calculation indicates that by enhancing
the impurity scattering, the frequency of the gap edge
will become independent of the relative directions of
the external fields ($A^x_{\omega}$ and ${\mib A}_0$).
By conducting such an experiment,
it will be possible
to make a relative comparison between
the impurity scattering and
the magnitude of electron--acoustic phonon interaction.

\section*{Acknowledgment}

The numerical computation in this work was carried out
at the Yukawa Institute Computer Facility.


\begin{thebibliography}{9}
\bibitem{note1}
Several studies on
strong-coupling superconductors~\cite{rainer,bickers}
show that 
there is a peak around $\omega\simeq 2\Delta$ in
the real part of the optical conductivity
even in the case of $e{\mib v}_{\mib k}\cdot{\mib A}_0=0$.
Then, it is possible that the weak-coupling superconductors,
in which the diagonal part of the self-energy by
the electron-phonon interaction 
is negligible, are suitable for the observation of changes in
the spectrum around the gap edge with
$e{\mib v}_{\mib k}\cdot{\mib A}_0\ne 0$.
\end{thebibliography}
\end{document}